\documentclass[review]{elsarticle}

\usepackage{hyperref}
\usepackage{graphicx}
\usepackage{dcolumn}
\usepackage{bm}
\usepackage{upgreek}
\usepackage{xcolor}
\usepackage{amsmath,amssymb}

\journal{International Journal of Multiphase Flow}

\biboptions{authoryear}

\begin{document}

\begin{frontmatter}

\title{Towards realistic simulations of human cough:\\effect of droplet emission duration and spread angle}

\author[a1,a2]{Mogeng Li}
\author[a8,a1,a2]{Kai Leong Chong} 
\author[a1,a2]{Chong Shen Ng}
\author[a3]{Prateek Bahl}
\author[a3]{Charitha M. de Silva}
\author[a1,a2,a4,a5]{Roberto Verzicco}
\author[a3]{Con Doolan}
\author[a6,a7]{C. Raina MacIntyre}
\author[a1,a2]{Detlef Lohse\corref{mycorrespondingauthor}}
\cortext[mycorrespondingauthor]{Corresponding author}
\ead{d.lohse@utwente.nl}

\address[a1]{Physics of Fluids Group, Max Planck Center for Complex Fluid Dynamics, J.\,M.\,Burgers Center for Fluid Dynamics and MESA+ Research Institute, Department of Science and Technology, University of Twente, 7500AE Enschede, The Netherlands}
\address[a2]{Max Planck Institute for Dynamics and Self-Organisation, 37077 G{\"o}ttingen, Germany}
\address[a8]{Shanghai Key Laboratory of Mechanics in Energy Engineering, Shanghai Institute of Applied Mathematics and Mechanics, School of Mechanics and Engineering Science, Shanghai University, Shanghai, 200072, PR China}
\address[a3]{School of Mechanical and Manufacturing Engineering, UNSW Sydney, Kensington, NSW, 2052, Australia}
\address[a4]{Dipartimento di Ingegneria Industriale, University of Rome `Tor Vergata', Roma 00133, Italy}
\address[a5]{Gran Sasso Science Institute - Viale F. Crispi, 7 67100 L'Aquila, Italy}
\address[a6]{Biosecurity Program, The Kirby Institute, UNSW Sydney, Kensington, NSW, 2052, Australia}
\address[a7]{College of Public Service \& Community Solutions and College of Health Solutions, Arizona State University, Phoenix, Arizona, USA}

\begin{abstract}
Human respiratory events, such as coughing and sneezing, play an important role in the host-to-host airborne transmission of diseases. Thus, there has been a substantial effort in understanding these processes: various analytical or numerical models have been developed to describe them, but their validity has not been fully assessed due to the difficulty of a direct comparison with real human exhalations. In this study, we report a unique comparison between datasets that have both detailed measurements of a real human cough using spirometer and particle tracking velocimetry, and direct numerical simulation at similar conditions. By examining the experimental data, we find that the injection velocity at the mouth is not uni-directional. Instead, the droplets are injected into various directions, with their trajectories forming a cone shape in space. Furthermore, we find that the period of droplet emissions is much shorter than that of the cough: experimental results indicate that the droplets with an initial diameter $\gtrsim\,10\upmu$m are emitted within the first 0.05\,s, whereas the cough duration is closer to 1\,s. These two features (the spread in the direction of injection velocity and the short duration of droplet emission) are incorporated into our direct numerical simulation, leading to an improved agreement with the experimental measurements. Thus, to have accurate representations of human expulsions in respiratory models, it is imperative to include parametrisation of these two features.
\end{abstract}


\end{frontmatter}

\section{\label{sec:intro}Introduction}

Since the outbreaks of SARS-CoV in 2003 and SARS-CoV-2 in 2019, the role played by the turbulent multiphase flow in the transmission of infectious diseases via the airborne route has received increasing attention. The host-to-host transmission of respiratory disease is a complicated process with multiple stages including the exhalation, dispersion and inhalation of the pathogen \citep{bourouiba2020turbulent, zhou2021dynamical, bourouiba2021fluid, chong2021extended, smith2020aerosol}.In particular, a central piece to the puzzle is the dispersion of pathogen-carrying droplets in turbulent flows.

A classic approach to mathematically predict the transmission of an infectious disease among populations is the compartmental models \citep{tolles2020modeling}. Developed in 1927, the most basic compartmental model is an SIR model \citep{kermack1927contribution}, where the population is separated into three so-called compartments, namely `Susceptible', `Infected' and `Removed'. Individuals can transfer between compartments based on ordinary differential equations while the total population remains constant. The infection rate of susceptible individuals can be improved with more physical insight via dose-response models \citep{brouwer2017dose}, or the Wells--Riley model in particular for diseases transmitted via the airborne route \citep{noakes2006modelling}. The application of these two groups of models and a detailed comparison are presented in the review paper by \citet{sze2010review}. In essence, the Wells--Riley model centers on the concept of `quantum', which is the dose required to start an infection, while the dose-response model uses the amount of pathogen taken in by the susceptible individual. In both approaches, the complicated transmission process that involves a series of stages (exhalation, dispersion, ventilation, inhalation, etc) is parametrised based on some physical assumptions or empirical observations \citep{lelieveld2020model, bazant2020beyond, jones2021modelling, nordsiek2021risk}. As a typical example, the room where the infectious disease is spread from is assumed to be `well-mixed', i.e. the pathogen distributes uniformly in the room. More recently, some efforts have been made to incorporate the temporal and spatial inhomogeneity in the model \citep{mittal2020mathematical,yang2020towards}. A sophisticated software tool has been developed to estimate the infection risk under a number of scenarios, including various modes of ventilation and different levels of activities \citep{mikszewski2020airborne}. These models are relatively easy to use and hence highly appealing when developing social guidelines based on the risk of transmission at a wide range of real-life scenarios. However, the derivation of these models, especially the selection of the parameters hinges on the detailed understanding of the governing physics in each step of the transmission. 

\begin{figure*}[hbt!]
\centering
\includegraphics[width=0.75\textwidth]{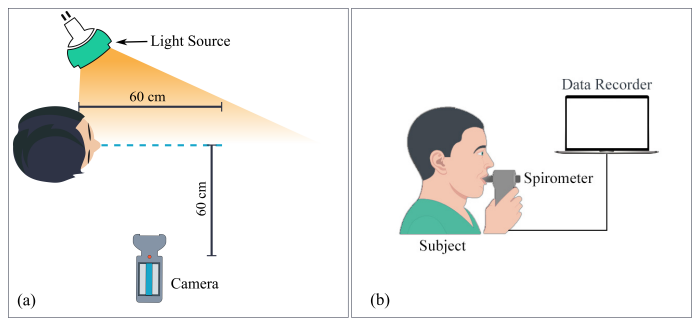}
\caption{\label{fig:exp_setup} (\textit{a}) Schematic of the setup used to capture high-speed frames of droplets expelled during coughing. (\textit{b}) Schematic of the spirometry test performed by the subject for coughing.}
\end{figure*} 

Alongside the effort in the epidemiology community, there has also been an ongoing endeavour in understanding and modelling the flow physics in the transmission of respiratory diseases. In the pioneering work of \citet{wells1934air}, the lifespan of a droplet is illustrated through a simple `evaporation-falling curve', where the fate of a small droplet is considered to be full evaporation, while that of a large droplet is ground deposition. This simplistic model was then improved by \citet{xie2007far}, where the salinity of the droplet, the ambient temperature and relative humidity conditions and the buoyant respiratory jet were taken into consideration. The effect of ambient conditions on the droplet evaporation is then incorporated into one parameter, the effective evaporation diffusivity proposed by  \citet{balachandar2020host}. The coupling between the carrying fluid and the disperse phase is further incorporated into the puff model by \citet{bourouiba_dehandschoewercker_bush_2014}, and a good agreement with the laboratory measurement of a multiphase puff is observed, but a direct comparison with clinical data, such as the speed and the trajectory of the puff is still lacking.

There is now a consensus that the flow physics in the transmission of respiratory diseases involve a \textit{multiphase flow} \citep{bourouiba2020turbulent}. Indeed, various human respiratory events, such as breathing, speaking, coughing and sneezing, can be simplified as transient turbulent jets with liquid droplets as the second phase \citep{balachandar2020host, mittal2020flow, bourouiba_dehandschoewercker_bush_2014, bourouiba2020turbulent, ng2020growth, chong2021extended, abkarian2020speech, stadnytskyi2020airborne}. The flow field associated with a single respiratory event have been investigated through both experiments with human subjects \citep{vansciver2011particle, bahl2020experimental} and laboratory or numerical simulations \citep{bourouiba_dehandschoewercker_bush_2014, wei2017human, chang2020aerial, abkarian2020speech, ng2020growth, chong2021extended, fabregat2021direct}. The general picture of a cough or sneeze is that the droplet-laden, warm and moist air is injected at a high speed over a short duration ($\lesssim 1$ s), which propagates under the effect of both the initial momentum and buoyancy \citep{bourouiba_dehandschoewercker_bush_2014,ng2020growth}, while in continuous events such as speaking, a `puff train' assimilates to a turbulent jet like structure in the far field \citep{abkarian2020speech}. The fate of the droplets varies largely depending on their size \citep{ng2020growth}: large droplets with a diameter $\sim\mathcal{O}(1000 \upmu\text{m})$ deposit on the ground in a near ballistic motion, while small droplets with a diameter $\sim\mathcal{O}(10 \upmu\text{m})$ are trapped in the humid puff and have prolonged lifetimes compared to the prediction of \citet{wells1934air}. Condensation is observed at low ambient temperature and high relative humidity settings \citep{chong2021extended}. Similar to the need for judicious assessment of the accuracy of theoretical models with clinical data, a direct comparison of numerical simulations with real respiratory events is imperative to test and elevate the accuracy of these models.

Accordingly, in this study, we report a unique pair of datasets of human coughs with both experimental and numerical components with similar conditions. The numerical dataset is obtained using direct numerical simulations (DNS) based on the methods employed by \citet{chong2021extended} and \citet{ng2020growth}, and the experimental component is the extension of the method used by \citet{bahl2020experimental} to coughs. We reveal that both the droplet emission duration and the initial spread angle have strong influence on the far-field behaviour of the flow field. These two refinements to the flow inlet conditions are validated through the experimental data collected from a human subject in \S\ref{sec:exp}. Two numerical simulation cases are formulated based on these refined conditions in \S\ref{sec:DNS}, and their detailed effect on the simulated flow is discussed in \S\ref{sec:results}.

\section{\label{sec:exp}Experimental observations}
To capture the flow dynamics of both the airflow and the droplets expelled during a cough from human exhalations, we utilise two sets of experiments. Namely, a spirometer is employed in the subject's mouth to capture the volumetric flow rate expelled during a cough (Fig.~\ref{fig:exp_setup}(\textit{b})), and Particle Tracking Velocimetry (PTV) experiments quantify the motion of the expelled droplets. For the latter, a volume illumination configuration is employed using a high powered pulsed LED light source (GsVitec L5), positioned at approximately $45^{\circ}$ from the image plane, to capture high-speed imaging data of the scattered light from the droplets expelled during a cough. The high-speed camera was placed $0.6$\,m away from the image plane and was equipped with a $35$\,~mm lens with an f-stop of F4 to capture a field-of-view of approximately $0.6\times0.35$\,m$^2$. This arrangement provided a resolution of 312.5\,$\mu$m/pixel and a depth of field of 40\,mm. Respiratory droplets expelled by the subject were used as tracer particles, and no additional seeding was introduced. The basic setup is depicted in Fig.~ \ref{fig:exp_setup}(\textit{a}). We note that all the experiments were performed at an ambient temperature of 22 $^{\circ}$C.

To perform precise particle tracking, the high-speed image sequences are first pre-processed to minimise background/sensor noise and isolate head movement. After that, PTV was performed on the image sequence using  Lavision\textregistered{} Davis~8.4. Here an initial Particle Image Velocimetry (PIV) pass on the image sequence was employed to obtain a velocity field estimate and then individual droplets were tracked using PTV algorithm \citep{Cowen1997}.  Further details on the experimental setup and processing can be found in \cite{bahl2020experimental}.

\begin{figure}
\centering
\includegraphics[width = 0.45\textwidth]{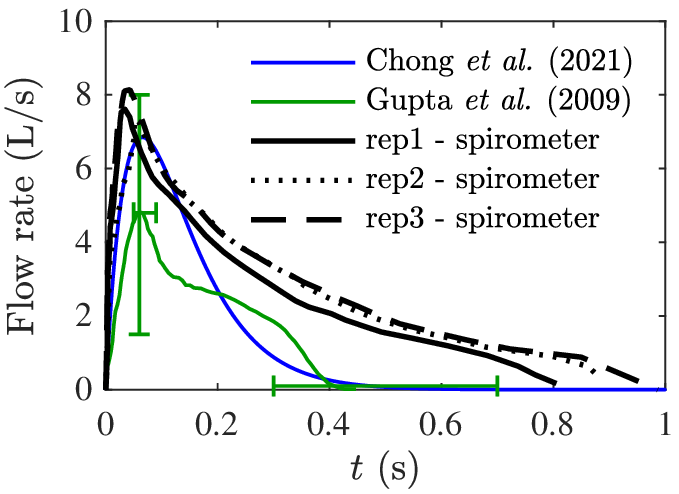}
\put(-170,110){(\textit{a})}
\hspace{10mm}
\includegraphics[width = 0.45\textwidth]{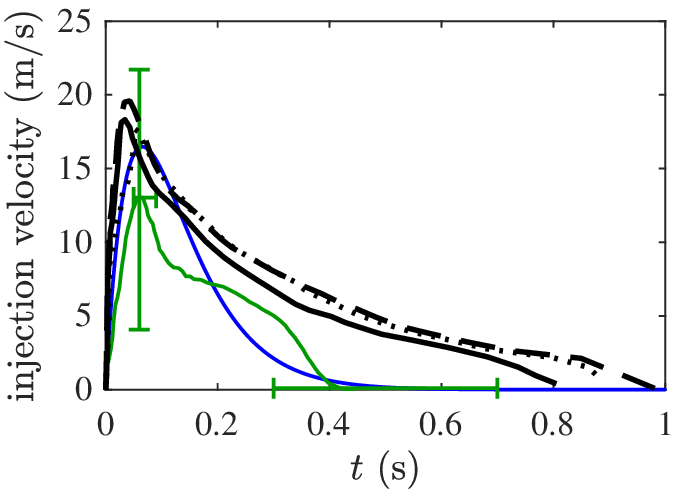}
\put(-175,110){(\textit{b})}
\caption{\label{fig:flow_rates} (\textit{a}) Flow rate and (\textit{b}) bulk injection velocity profile computed from the flow rates as a function of time. In the present study and \cite{gupta2009flow}, the flow rate is measured experimentally by a spirometer, whereas the inlet velocity profile used by \cite{chong2021extended} is converted to flow rate by assuming a circular mouth shape with a constant diameter $D = 23$ mm. The scatter of the peak location and magnitude from 25 subjects of \cite{gupta2009flow} is shown by the horizontal and vertical error bars at the peak flow rate, and the variation of the cough duration is shown by the horizontal error bar centred around $t = 0.5$ s. The three black curves (rep1, rep2 and rep3) are three repeating measurements of the same subject.}
\end{figure}

In order to draw a direct comparison of the flow rates in our experiments, Fig.~\ref{fig:flow_rates} shows the flow rate of coughs measured by the spirometer compared to the representative profile obtained using the same method in the experiments of \cite{gupta2009flow}, and the inlet velocity condition assumed in our previous work \citep{chong2021extended}. Although the flow rate of the experiments peaks within a shorter time after the start of the cough, the overall duration and intensity of the coughs are very comparable with previous works, i.e. the cough duration is $\mathcal{O}(1)$\,s. Three repeating measurements are conducted on the same subject, and an average of these measurements is used as the injection flow rate in the simulation to be detailed in \S\ref{sec:DNS}.

\begin{figure}
\centering
\includegraphics{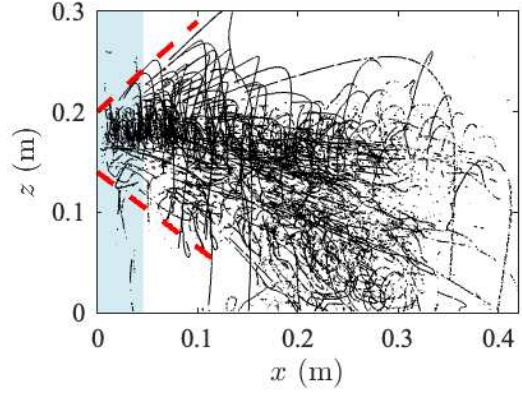}
\caption{\label{fig:Exp2_tracking} Droplet trajectories of a single cough captured by PTV. The flow direction is from left to right, and the mouth is located at $x = 0$ and at height $z\approx0.18$ m. The region marked by a blue shade has a streamwise extent of $0<x<2D$, where $D$ is the diameter of the mouth.}
\end{figure}

Fig.~\ref{fig:Exp2_tracking} shows the droplet trajectories of a single cough produced by the same subject and captured by PTV for a duration of approximately 0.4\,s. Note that as limited by the intrusive nature of the spirometer measurement, the trajectories presented here are not obtained simultaneously with the flow rate date shown in Fig.~\ref{fig:flow_rates}. However, the close agreement between the three repeating spirometer curves in Fig.~\ref{fig:flow_rates} indicates that measurements performed at different times by the two methods can still be reliably compared. The droplets are emitted into various directions, and the ballistic trajectories form a cone shape in the space with an apex angle around $90^{\circ}$, as delineated by the red dashed lines. Such spreading phenomena can also be observed in the droplet trajectories captured by \cite{bourouiba_dehandschoewercker_bush_2014} and the inline holography in \cite{shao2021risk}. In the atomisation process of a high-speed gas jet shearing a liquid film, which is considered as a major mechanism of droplet generation in respiratory tracts \citep{johnson2009mechanism}, there is also experimental evidence that droplets are ejected at an angle of $20^{\circ}-40^{\circ}$ relative to the jet flow direction \citep{descamps2008gas}. Note that the spreading angle of the initial velocity field discussed here is a near field quantity ($\mathcal{O}(D)$ where $D$ is the mouth diameter), which is different from the far field ($>\mathcal{O}(5D)$) spreading angle of the respiratory jet typically reported to be $ 10^{\circ} - 15^{\circ}$ with respect to the centre-line \citep{gupta2009flow, abkarian2020speech, dudalski2020experimental}.

\begin{figure}
\centering
\includegraphics{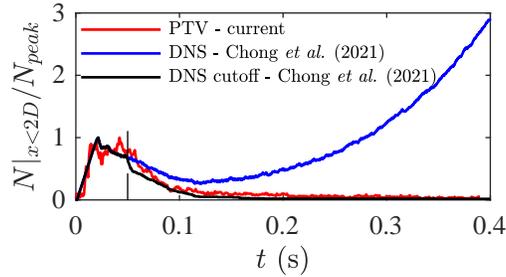}
\caption{\label{fig:Exp_DNS_droplet_num} Number of droplets within $x<2D$ from the mouth in each frame, normalised by $N_{peak}$, magnitude of the first peak in the $N|_{x<2D}$ versus $t$ curve. The red curve is from the present experimental (PTV) dataset, while the blue curve is from the DNS results of \cite{chong2021extended}. The black curve is from the same simulation as the blue curve, but with droplet emission suppressed after $t_{\text{cutoff}} = 0.05$ s, which is marked by the vertical dashed line in the figure. }
\end{figure}
By visually inspecting the raw PTV images, the droplet emission rate is found to be the highest at the beginning of a cough, then it quickly diminishes long before the flow injection velocity at the mouth reaches zero \citep{DeSilva2021}. In other words, the droplet emission is concentrated at the onset of the cough for a much shorter duration than that of the flow injection. To provide further quantitative evidence regarding this observation, we utilise the numerical dataset in \cite{chong2021extended}, and compare the number of droplets in a region close to the mouth (as marked by the blue shaded box in Fig.~\ref{fig:Exp2_tracking}) before and after removing all droplets emitted after $t_{\text{cutoff}} = 0.05$\,s. We choose to focus on this region mostly due to the consideration that here the droplet number is predominantly dependent on the emission and streamwise transportation, with negligible contribution from other factors such as evaporation, deposition or droplets leaving the measurement volume. Fig.~\ref{fig:Exp_DNS_droplet_num} shows the number of droplets within $2D$ from the mouth in the streamwise direction in each frame, where $D$ is the mouth diameter. Each curve is normalised by the magnitude of the first peak emerging around $t\approx0.03$\,s, thus eliminating the effect of the total number of droplets on the comparison between cases. In the PTV dataset, the number of droplets captured in a frame reaches its maximum between 0.02\,s and 0.04\,s, before it approaches 0 at 0.15\,s after the start of the cough, when the injecting flow rate is still more than 60\% of its peak value (see Fig.~\ref{fig:flow_rates}). The DNS case of \cite{chong2021extended} shows a similar trend as in the PTV data between 0 and 0.1\,s, but instead of reducing to 0, the droplet number then quickly increases and the value almost triples at 0.4\,s, indicating a very dense droplet population close to the mouth. In the DNS data of \cite{chong2021extended}, droplets were emitted at a constant rate over a duration of 1\,s. At the later stage of the cough when the injection velocity was low, the droplets still entered the flow field at a constant rate, leading to the accumulation of droplets close to the mouth. After removing all the droplets emitted after $t_{\text{cutoff}} = 0.05$\,s (black curve in Fig.~\ref{fig:Exp_DNS_droplet_num}), better agreement with the PTV result is observed. 

Note that the PTV method in this study can confidently detect droplets with $d\gtrsim10\upmu $m although smaller droplets are also likely to be detected. Therefore, the emission cutoff time discussed above is only strictly applicable to droplets in this size range, and it is likely that smaller aerosol particles formed deep in the lung \citep{johnson2009mechanism} are emitted throughout the entire cough when the air flow is present.

To summarise, two major features of a real human cough can be concluded from the experiments: first, the initial velocities of droplets are not uni-directional. Instead, droplets with a relatively high momentum spread into various directions after exiting the mouth, as evidenced by their trajectories which form a cone shape with an apex around $90^{\circ}$. Second, the majority of droplets with $d\gtrsim10 \upmu$m are emitted between 0-0.05\,s, which is less than 10\% of the duration of a cough based on a nonzero injection flow rate. Incorporating these features in numerical simulations are therefore mandatory in order to achieve a better representation of realistic human respiratory events.

\begin{figure}
\centering
\includegraphics[scale = 0.3, trim = 40mm 180mm 180mm 40mm, clip]{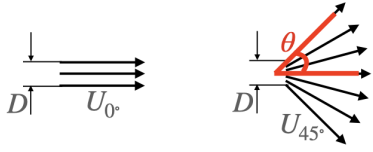}
\caption{\label{fig:inlet_config} Schematics of the flow inlet conditions in DNS. The left figure is for the $\theta = 0^{\circ}$ (uni-direction) case, and the right one is for $\theta = 45^{\circ}$. }
\end{figure}

\section{\label{sec:DNS}Numerical investigations}

In this section, we provide details of the numerical cases designed to study the effect of the spread angle and droplet emission cutoff time. 

We consider an incompressible fluid of gas phase, with both temperature and vapour concentrations coupled to the velocity field by employing the Boussinesq approximation. The gas phase is solved using DNS by a staggered second-order accurate finite difference scheme and marched in time using a fractional-step third-order Runge--Kutta approach \citep{verzicco1996finite,vanderpoel2015}. Buoyancy effect of the puff as a result of the temperature and vapour mass fraction difference between the exhalation and the ambient environment is considered in the simulation. For droplets, we apply the spherical point-particle model, and consider the conservation of momentum (Maxey-Riley equation), energy, and mass. During the evaporation of droplets, the temperature and vapour mass fraction of the surrounding air are modified, which leads to a change in the density of the local gas phase. The momentum equation of the gas phase is not affected by the presence of droplets except through the modification of density. We use the measurement of \cite{xie2009exhaled} for the initial droplet size distribution. For the domain size and other details on the implementation, we refer readers to \cite{chong2021extended} and \cite{ng2020growth}.

The spread angle, $\theta$, is defined as half of the cone apex angle. Therefore the $90^{\circ}$ apex angle observed in Fig.~\ref{fig:Exp2_tracking} is equivalent to $\theta = 45^{\circ}$. Two configurations are simulated in the present numerical study, with $\theta = 0^{\circ}$ and $\theta = 45^{\circ}$, respectively. Schematics of the inlet conditions can be found in Fig.~\ref{fig:inlet_config}. The initial velocity vectors at the inlet cross section have the same magnitude within each case, and they spread radially in the $\theta = 45^{\circ}$ case. The velocity distribution is then smoothed with an axis-symmetric Gaussian function.

The mouth shape in both cases is assumed to be a circle with diameter $D = 23$\,mm, which remains constant in time. By integrating the streamwise velocity across the mouth area, the flow rate of the exhalation can then be expressed as  
\begin{subequations}
    \begin{equation}
    \dot{m}_{0^{\circ}} = \rho \int_0^{D/2}2\pi rG(r)U_{0^{\circ}}\mathrm{d}r
    \end{equation}
    and
    \begin{equation}
    \dot{m}_{45^{\circ}} =\rho\int_0^{D/2}2\pi r
G(r)U_{45^{\circ}}\cos{\phi(r)}\mathrm{d}r,
    \end{equation}
    \label{eq:flow_rate}
\end{subequations}
where $\rho$ is the air density, $G(r)$ is the Gaussian function and $r$ is the variable of integration with $r = 0$ at the centre-line of the mouth. $\phi(r)$ is the angle between the injection velocity vector and the horizontal axis in the $\theta = 45^{\circ}$ case, which can be related to $r$ as
\begin{equation}
    \tan \phi(r) = \frac{r}{D/2}.
\end{equation}

The injected flow rate and forward momentum govern the propagation of the respiratory jet. For a given flow rate, the injection velocity magnitudes, $U_{0^{\circ}}$ and $U_{45^{\circ}}$, can then be determined using (\ref{eq:flow_rate}). Here we assume that both cases have the same flow rate profile as the spirometer result in the present study (see Fig.~\ref{fig:flow_rates}). The resulting centreline velocity of the $\theta = 45^{\circ}$ case is approximately 16\% higher compared to the $\theta = 0^{\circ}$ case. The injected forward momentum of the former is then 7\% higher than the latter.

\begin{figure*}[t!]
\includegraphics[width = \textwidth]{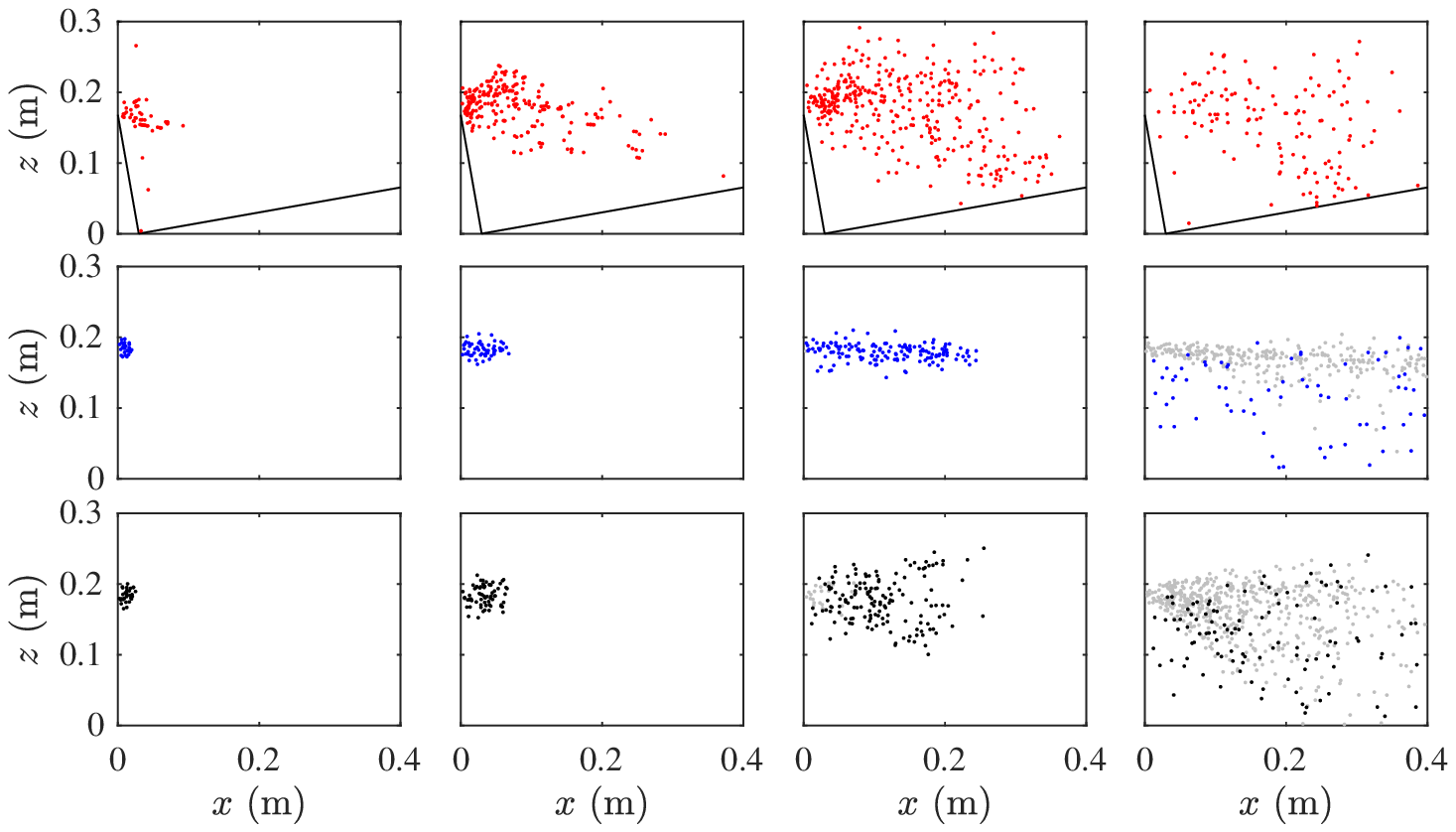}
\put(-318,183){(\textit{a})}
\put(-236,183){(\textit{b})}
\put(-154,183){(\textit{c})}
\put(-72,183){(\textit{d})}
\put(-318,124){(\textit{e})}
\put(-236,124){(\textit{f})}
\put(-154,124){(\textit{g})}
\put(-72,124){(\textit{h})}
\put(-318,65){(\textit{i})}
\put(-236,65){(\textit{j})}
\put(-154,65){(\textit{k})}
\put(-72,65){(\textit{l})}
\put(-375,165){Exp.}
\put(-375,106){DNS}
\put(-376,97){$\theta = 0^{\circ}$}
\put(-375,50){DNS}
\put(-378,41){$\theta = 45^{\circ}$}
\put(-308,200){$t = 0.01$ s}
\put(-226,200){$t = 0.02$ s}
\put(-144,200){$t = 0.05$ s}
\put(-62,200){$t = 0.20$ s}

\caption{\label{fig:time_sequence} Droplet distribution of (\textit{a-d}) the experiment, (\textit{e-h}) DNS with a spread angle of $\theta = 0^{\circ}$ and (\textit{i-l}) $\theta = 45^{\circ}$. From left to right, the four columns correspond to $t = 0.01$\,s, 0.02\,s, 0.05\,s, and 0.20\,s. The experimental case is rotated by $10^{\circ}$ counter-clockwise to align the cough with the streamwise direction for better comparison, and the original field-of-view is marked by the solid lines. Only droplets with an initial diameter greater than $10 \,\upmu$m are shown in the DNS results, and only every the second droplet is shown in the figure. The grey dots in (\textit{h}) and (\textit{l}) are droplets emitted between 0.05\,s to 0.20\,s. }
\end{figure*}

A total number of 5000 droplets with the initial size distribution following \cite{xie2009exhaled} are introduced to the simulation domain at a constant rate over 0.4\,s. The total number of droplets is also chosen based on the same study. Since the droplet distribution in the flow field is relatively sparse, the droplet-to-droplet interaction is negligible. For the same reason the modification to evaporation rates due to droplet clustering is also insignificant. Therefore, it is possible to remove droplets emitted after a certain time in post processing to study the effect of the droplet emission cutoff time.

\section{\label{sec:results}Results and discussion}
The spatial distribution of the droplets in the DNS cases at a range of time instances from 0.01 s to 0.2 s are shown in Fig.~\ref{fig:time_sequence}. Overall, the $45^{\circ}$ spread angle case qualitatively agrees with the experimental results: droplets form a `cloud' after entering the flow field, instead of concentrating along the centre-line as in the $0^{\circ}$ case. Both the experiment and the $\theta = 45^{\circ}$ case have a large population of droplets above the centre-line of the mouth ($z_m = 0.184$ m), which is absent in the $\theta = 0^{\circ}$ case. Furthermore, the droplet density near the mouth is much lower in Fig.~\ref{fig:time_sequence}(\textit{d}) compared to earlier time instances, and such a decrease is also observed in the simulation in Fig.~\ref{fig:time_sequence}(\textit{l}) after excluding all the grey dots, which are droplets emitted after 0.05 s. This observation confirms our previous hypothesis that the rate of droplet emission during a cough varies with time, and large droplets are released only at the early stage of the cough. In the literature, the common practice is to assume that the droplets are uniformly distributed in the discharging fluid \citep{bourouiba_dehandschoewercker_bush_2014, wei2017human} or emitted at a constant rate over the entire duration of the cough \citep{chong2021extended, ng2020growth, fabregat2021direct}. Incorporating the short droplet emission time in the model can significantly improve the agreement between the simulation and the experiment.  

\begin{figure*}
\vspace{1mm}
\includegraphics[width = \textwidth]{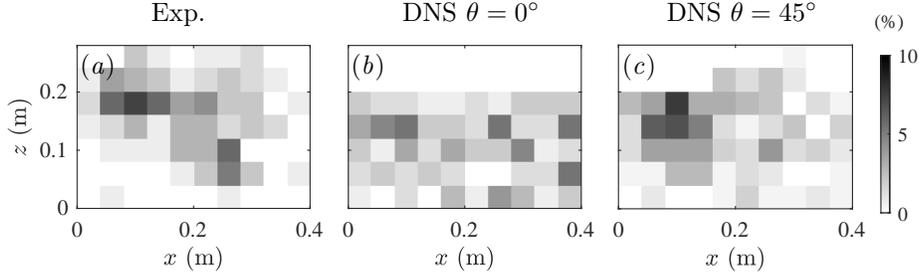}
\put(-317,75){(\textit{a})}
\put(-214.5,75){(\textit{b})}
\put(-112,75){(\textit{c})}
\put(-290,95){Exp.}
\put(-195,95){DNS $\theta = 0^{\circ}$}
\put(-95,95){DNS $\theta = 45^{\circ}$}

\caption{\label{fig:hist_compare} Histogram of droplet distribution at $t = 0.20 $ s of (\textit{a}) the experiment, (\textit{b}) DNS with a spread angle of $\theta = 0^{\circ}$, and (\textit{c}) $\theta = 45^{\circ}$. Only droplets emitted between 0 and 0.05 s are included in the histogram.}
\end{figure*}

The contribution of the spread angle is further analysed in Fig.~\ref{fig:hist_compare}, a 2D histogram of the droplet spatial distribution at $t = 0.20$ s for the experiment and the $\theta = 0^{\circ}$ and $\theta = 45^{\circ}$ DNS cases. The short droplet emission time is taken into account, thus only droplets emitted before $t = 0.05$ s are included when computing the histogram. The histogram peaks at $x \approx 0.1$ m and $z\approx 0.19$ m in the experiment (Fig.~\ref{fig:hist_compare}\textit{a}), with droplets scattered both above and below $z_m$. These major features are successfully replicated in the $\theta = 45^{\circ}$ case (Fig.~\ref{fig:hist_compare}\textit{c}). In the $\theta = 0^{\circ}$ case (Fig.~\ref{fig:hist_compare}\textit{b}), however, very few droplets can be found above the mouth location $z_m$. The build-up of droplets near the mouth is absent, and the droplet distribution is almost uniform in the streamwise direction.

\begin{figure*}[t!]
\includegraphics[width = 0.495\textwidth]{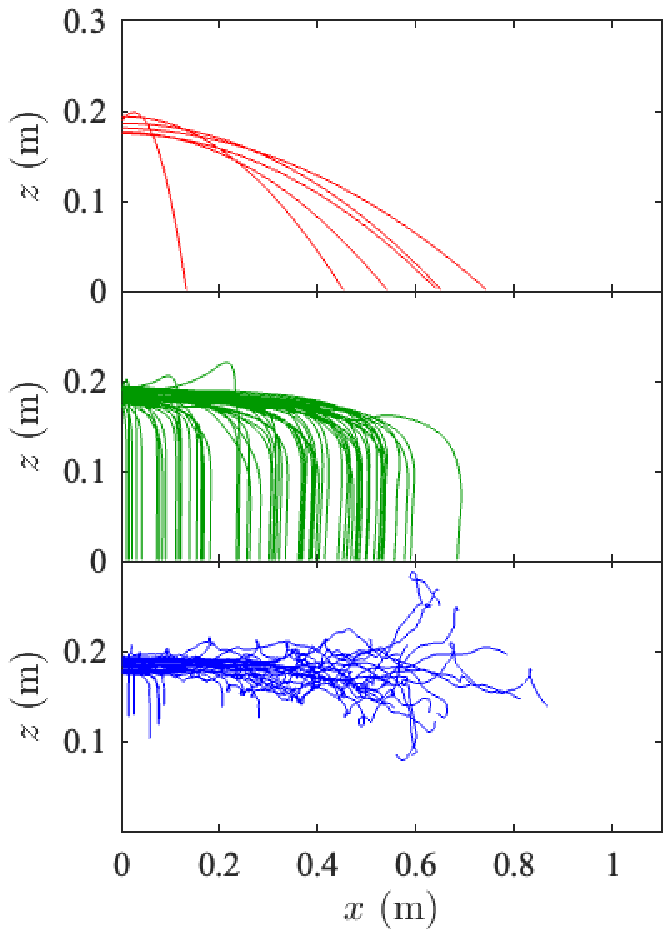}
\includegraphics[width = 0.495\textwidth]{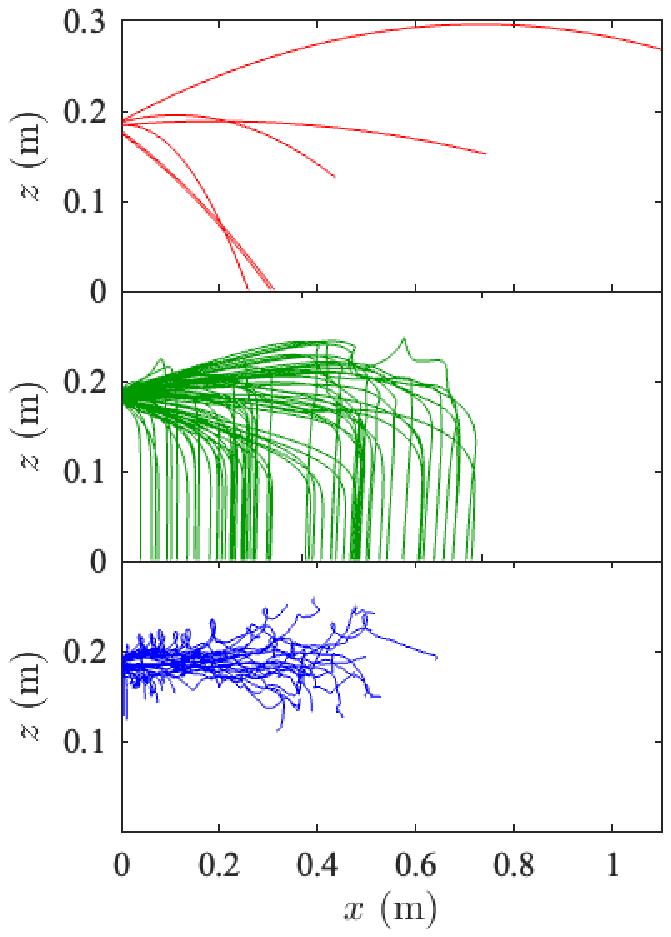}
\put(-314,208){(\textit{a})}
\put(-140,208){(\textit{b})}
\put(-314,147){(\textit{c})}
\put(-140,147){(\textit{d})}
\put(-314,86){(\textit{e})}
\put(-140,86){(\textit{f})}
\put(-390,183){Large}
\put(-390,122){Medium}
\put(-390,61){Small}
\caption{\label{fig:DNS_traj_cone} Droplet trajectories at $t = 0-0.2$\,s for DNS cases $\theta = 0^{\circ}$ (left column) and $\theta = 45^{\circ}$ (right column). The top, middle and bottom row correspond to large, medium and small droplets, with an initial diameter of $d_0>500\,\upmu$m, $d_0\approx90\,\upmu$m and $d_0<40\,\upmu$m. All the trajectories shown in the figure are from droplets emitted within $t = 0-0.05$\,s. }
\end{figure*}

Droplet trajectories of cases $\theta = 0^{\circ}$ and $\theta = 45^{\circ}$ are compared side-by-side in Fig.~\ref{fig:DNS_traj_cone}. As expected, droplets with an initial diameter $d_0>500\,\upmu$m exhibit predominately ballistic behaviour, and a non-zero spread angle introduces a larger scatter in their trajectories.
For the medium-sized droplets, the streamwise deceleration is complete before they reach the floor. Comparing Fig.~\ref{fig:DNS_traj_cone}(\textit{c}) and (\textit{d}), one would notice that droplets in the $\theta = 45^{\circ}$ case settle closer to the mouth. Instead of carried downstream by the jet with a similar velocity, droplets emitted with a large spread angle penetrate deeper sideways into the still ambient fluid, so they experience larger relative velocity (therefore higher aerodynamic drag and stronger deceleration) compared to the $\theta = 0^{\circ}$ case. The reduction in the settling distance is not limited to droplets emitted with a high wall-normal velocity component: it is also seen in those with small or zero wall-normal component, as a result of the lower streamwise velocity in the widened jet.

\begin{figure}
\centering
\includegraphics{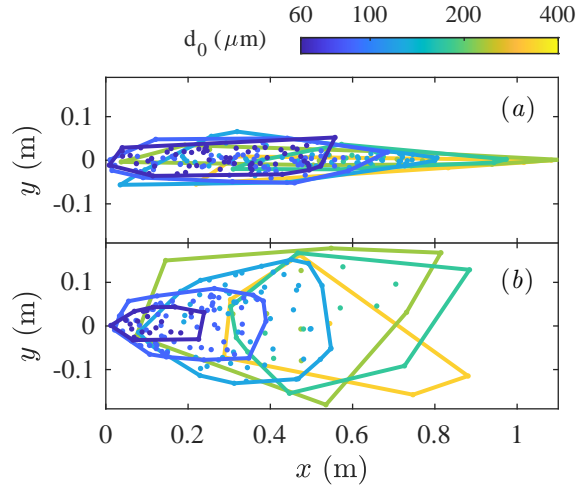}
\put(-30,140){(\textit{a})}
\put(-30,75){(\textit{b})}
\vspace{4mm}
\caption{\label{fig:group_deposition} Top-view of the droplet deposition location on the ground for (\textit{a}) $\theta = 0^{\circ}$ and (\textit{b}) $\theta = 45^{\circ}$ cases. The colour of each dot is determined by the droplet initial diameter $d_0$, which changes from blue to yellow with increasing $d_0$. The solid lines are the convex hull of the deposition locations of droplets with the same $d_0$.}
\end{figure}

The settling behaviour of droplets is further explored by examining the distribution of deposition locations in the ground plane of $z=0$ in Fig.~\ref{fig:group_deposition}. The location where a droplet reaches the ground is represented by a dot in the $x-y$ plane, with the colour determined by its initial diameter $d_0$. The contaminated region on the ground for each size class is shown by the convex hull. Although the mouth is at $z\approx 0.18$ m in the simulation domain, for smaller droplets ($d_0\lesssim100 \,\upmu$m), the streamwise velocity has already decreased to 0 before reaching the $z=0$ plane. Therefore, the deposition pattern reported in Fig.~\ref{fig:group_deposition} would still be applicable when the mouth is at $z\approx1.5$ m above the ground, which is the case of a standing person.

\begin{figure}
\centering
\includegraphics{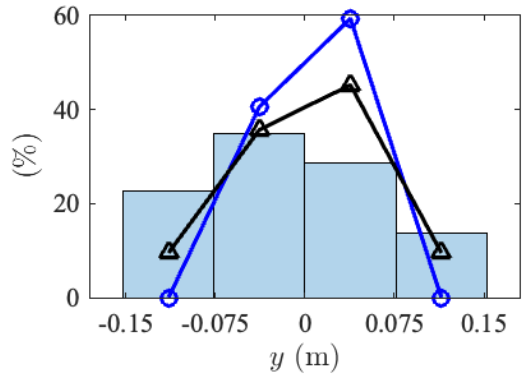}
\caption{\label{fig:group_deposition_barchart} Number percentages of deposited droplets in four spanwise bins at $x\approx 0.2$ m. The $x$-axis labels indicate the edges of each bin. The light blue bars are data from \cite{xie2009exhaled}, while the blue and black lines represent cases $\theta = 0^{\circ}$ and $\theta = 45^{\circ}$, respectively.}
\end{figure}

Comparing Fig.~\ref{fig:group_deposition}(\textit{a}) and (\textit{b}), it is apparent that the deposition location scatters over a wider range in the spanwise direction for $\theta = 45^{\circ}$, while it is mostly concentrated along the centre-line in the $\theta = 0^{\circ}$ case. This is a progress towards a more realistic cough, as evidenced by the improved agreement with the data of human subjects \citep{xie2009exhaled} (Fig.~\ref{fig:group_deposition_barchart}). In Fig.~\ref{fig:group_deposition}(\textit{b}), the spanwise width of the convex hull increases with increasing $d_0$, presumably because larger droplets have a higher initial momentum and they can continue for a longer distance along the direction of their initial velocity. The streamwise extent of the convex hull is shorter in the $\theta = 45^{\circ}$ case compared to the $\theta = 0^{\circ}$ case, indicating that droplets (especially those with $d_0\lesssim100\,\upmu$m) have a strong tendency to deposit closer to the mouth. 

We note that the initial size distribution of droplets exhaled during respiratory events is presently actively investigated. The simulation results can be affected by the size distribution: for instance, if we use the distribution of \cite{duguid1946}, which has more small ($\mathcal{O}(10)$\,$\upmu$m) droplets, a higher number of trajectories can be expected in Fig.~\ref{fig:DNS_traj_cone}(\textit{e-f}). However, here our focus is on capturing the dominant effects of an actual cough.

\begin{figure*}
\centering
\setlength{\unitlength}{1cm}
\begin{picture}(12.4,7)
\put(0,3.5){\includegraphics[scale = 0.85]{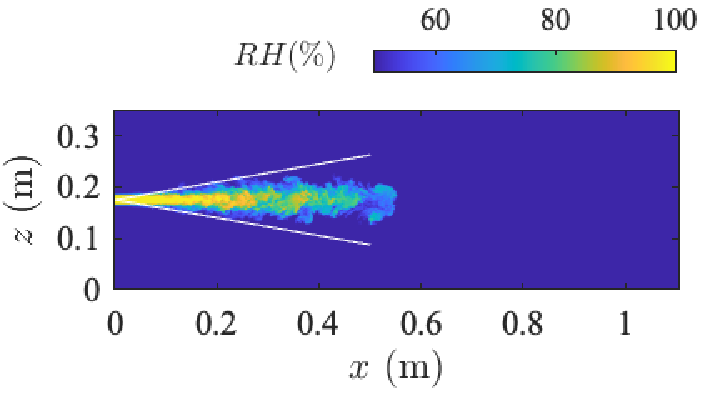}}
\put(0,0){\includegraphics[scale = 0.85]{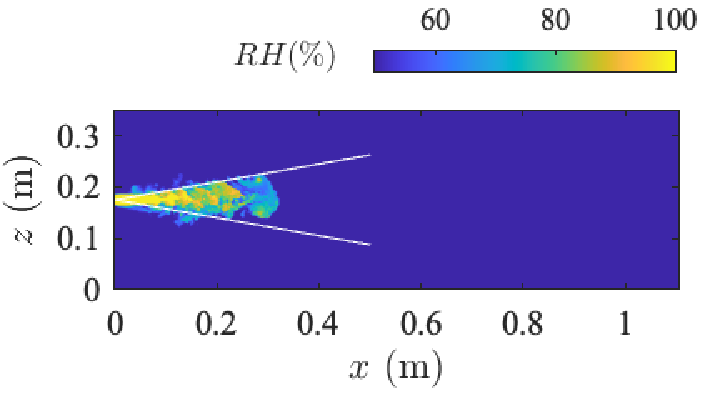}}
\put(6.16,3.5){\includegraphics[scale = 0.85]{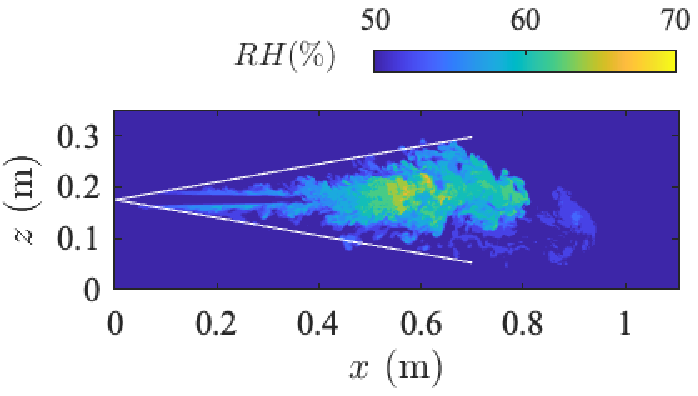}}
\put(6.16,0){\includegraphics[scale = 0.85]{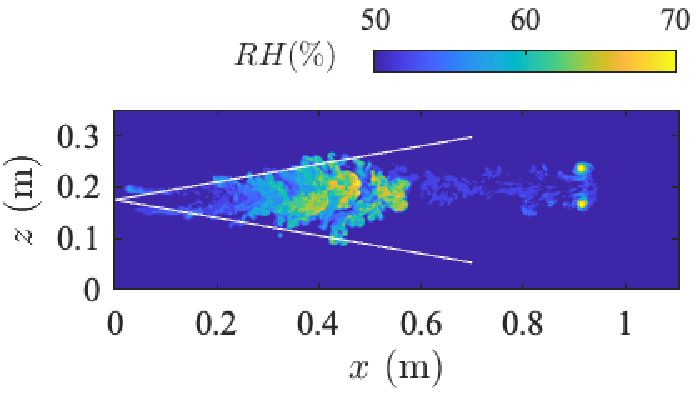}}
\put(0.1,6.8){(\textit{a})}
\put(6.26,6.8){(\textit{b})}
\put(0.1,3.4){(\textit{c})}
\put(6.26,3.4){(\textit{d})}
\end{picture}
\caption{\label{fig:RH_field} Relative humidity in the spanwise midplane ($y=0$) for (\textit{a},\textit{c}) $\theta = 0^{\circ}$ and (\textit{b},\textit{d}) $\theta = 45^{\circ}$ cases. The top row is at $t_1 = 0.15$ s and the bottom row is $t_2 = 0.6$ s. The white lines in each plot form a cone angle of $\alpha = 10^{\circ}$.}
\end{figure*}

The scalar fields of relative humidity ($RH$) are shown in Fig.~\ref{fig:RH_field}. At $t_1 = 0.15$~s, despite the higher injected streamwise momentum, the $\theta = 45^{\circ}$ case has a shorter and wider turbulent jet compared to the $\theta = 0^{\circ}$ case. The flow also appears more turbulent close to the mouth ($x\lesssim0.2$ m) in the former. The cone angle $\alpha$ of the jet, which can be determined from $\tan\alpha = R_{\text{jet}}(x)/x$, where $R_{\text{jet}}(x)$ is the jet radius at the streamwise location $x$, is very different from the spread angle $\theta$. Comparing the white reference lines in the figure with the scalar field of the jet, it is shown that the jet cone angle is approximately $10^{\circ}$ in the $\theta = 45^{\circ}$ case, while it is only slightly smaller in the $\theta = 0^{\circ}$ case. Therefore, a $45^{\circ}$ change in the spread angle $\theta$ only results in a less than $5^{\circ}$ change in the cone angle $\alpha$.

At $t_2 = 0.6$ s (Fig.~\ref{fig:RH_field}\textit{c} and \textit{d}), a vortex ring separated from the main jet is observed in both cases. These vortices are formed at the front of the jet, which can be seen attached to the jet at $t = t_1$. As the injection velocity decreases and the jet front decelerates, the self-induced velocity of the vortex ring eventually exceeds the velocity of the jet front, then the vortex ring detaches from the jet. Such vortex rings have also been observed in the large-eddy simulation results by \cite{liu2021peering}, and their formation, strength and velocity are shown to be very sensitive to the initial conditions of the simulation.

\section{\label{sec:conclusion}Conclusions and outlook}
In this study, we present a unique comparison of both experimental and numerical datasets on a human cough with matched conditions. Specifically, we implemented two important features of the human cough with insight gained from the comparison with the experimental data: the injection of the coughing flow which is shown to be better described with a spread angle of $45^{\circ}$, and the emission of large droplets only occurs for a short period of time ($\approx 0.05$ s) at the beginning of a cough, whose length is much longer ($\approx 1$ s). The contribution of these features to the simulation result is examined by comparing two DNS cases with the conventional and with the modified inlet conditions based on the aforementioned experimental observations from a human cough. Overall, a significant improvement in the agreement with the experimental data is observed after implementing the modifications. We further demonstrate that by considering the $45^{\circ}$ initial spread angle, the simulated jet appears thicker and shorter. Droplets deposit on the ground closer to the mouth in the streamwise direction, but scatter over a wider range in the spanwise direction, which is consistent with the experimental observations. The aim of future work must be to extend the present comparative approach between experiments and numerical simulations to other respiratory events such as singing and in particular speaking, which accumulatively plays an even more important role in the aerosol release \citep{abkarian2020speech,yang2020towards}.

\section*{Acknowledgements}
This work was funded by the Netherlands Organisation for Health Research and Development (ZonMW), project number 10430012010022: ``Measuring, understanding \& reducing respiratory droplet spreading'', the ERC Advanced Grant DDD, Number 740479, Foundation for Fundamental Research on Matter with Project No. 16DDS001, which is financially supported by the Netherlands Organisation for Scientific Research (NWO). C de Silva acknowledges ARC LIEF funding Grant, Number LE200100042, which supported the experimental work. The use of the national computer facilities in this research was subsidized by NWO Domain Science. We also acknowledge PRACE for awarding us access to MareNostrum in Spain at the Barcelona Computing Center (BSC) under the project 2020235589 and Irene at Tr\`{e}s Grand Centre de calcul du CEA (TGCC) under PRACE project 2019215098. KL Chong is also supported by Shanghai Science and Technology Program under the project no. 19JC1412802. 

\section*{Data availability}
As part of the ZonMW funding agreement, data of the DNS cases are available as open access, which can be accessed via the link \url{https://doi.org/10.4121/17099468}.  Larger files such as the scalar fields will be available upon request. The experimental data cannot be directly shared due to ethics limitations.

\bibliography{apssamp}

\end{document}


\baselineskip24pt
	\maketitle 

\section*{Governing equations} \label{sec:GovEqns}

The governing equations include the equations for gas phase and equations for droplets \cite{russo2014}. Both temperature and vapour mass fraction (note that the vapour mass fraction is defined as concentration of water vapour over the density of the gas) are coupled to the velocity field by employing the Boussinesq approximation. The motion of the gas phase is assumed to be incompressible $\partial {u}_i / \partial {x}_i = 0$, and governed by the momentum, energy, and vapour mass fraction equations:
\begin{equation}
    \frac{\partial{u_i}}{\partial t} + u_j \frac{\partial{u_i}}{\partial x_j} = -\frac{\partial p}{\partial x_i} + \nu_{air}\frac{\partial^2{u_i}}{\partial x_j^2} + g (\beta_\theta \theta + \beta_c c)\hat{e}_y,
    \label{eqn:GasMomen}
\end{equation}
\begin{equation}
    \rho_g c_{p,g} (\frac{\partial \theta_g}{\partial t} + u_i\frac{\partial \theta_g}{\partial x_i})= k_g \frac{\partial^2 \theta_g}{\partial x_i^2} - \sum_{n=1}^N c_{p,g}\theta_{g,n} \frac{\mathrm{d} m_n}{\mathrm{d} t} \delta(\vec{x}-\vec{x}_n) 
    - \sum_{n=1}^N h_m A_n (\theta_{g,n}-\theta_n) \delta(\vec{x}-\vec{x}_n),
    \label{eqn:GasTemp}
\end{equation}
\begin{equation}
\frac{\partial{c}}{\partial{t}} + u_i \frac{\partial c}{\partial x_i} = D_{vap}\frac{\partial^2 {c}}{\partial{x}_i^2} - \sum_{n=1}^N \left( \frac{\rho_l}{\rho_g} {A}_n \frac{\mathrm{d} {r_n}}{\mathrm{d}{t}} \delta(\vec{x}-\vec{x}_n) \right).
\label{eqn:GasMass}
\end{equation}
The last term on the right-hand side of \equrefs{eqn:GasMomen} represents the coupling of temperature and vapour mass fraction to the momentum equation.

For droplets, the spherical point-particle model is applied, and we consider the conservation of momentum (Maxey-Riley equation \cite{maxey1983}), energy, and mass as follows:

\begin{equation}
    \frac{\mathrm{d} u_{i,n}}{\mathrm{d} t} = (\beta+1) \frac{\mathrm{D} u_{i,g,n}}{\mathrm{D} t} + (\beta+1) \frac{3 \nu_{air} (u_{i,g,n}-u_{i,n})}{r_n^2} + g\beta \hat{e}_y,
    \label{eqn:DropletMomen}
\end{equation}
\begin{equation}
    \rho_l c_{p,l}V_n\frac{\mathrm{d}\theta_n}{\mathrm{d}t} = \rho_l A_n L\frac{\mathrm{d}r_n}{\mathrm{d}t} + h_m A_n(\theta_{g,n}-\theta_n),
    \label{eqn:DropletTemp}
\end{equation}
\begin{equation}
    \frac{\mathrm{d}r_n}{\mathrm{d}t} = -\frac{D_{vap} Sh_{drop}}{2 r_n}\frac{\rho_g}{\rho_l}\ln \left(\frac{1-c_{fluid}}{1-c_{drop}}\right).
    \label{eqn:DropletDiam}
\end{equation}

The notations we used here are as follows: ${u}_i$, ${u}_{i,n}$, and ${u}_{i,g,n}$ are the velocity of gas, droplets, and gas at the location of droplets, respectively. Similarly ${\theta}_g$, ${\theta}_n$, and ${\theta}_{g,n}$ are in Kelvin and used to represent the temperature of gas, droplets, and gas at the location of droplets, respectively. $c$ is the vapour mass fraction, ${r_n}$ the droplet radius, ${A}_n$ the surface area of the droplets, ${V}_n$ the volume of the droplets, $m_n$ the mass of the droplets. Also ${p}$ denotes the reduced pressure. $h_m$ is the heat transfer coefficient. $L$ is the latent heat of vaporisation of the liquid. $\rho_l$ and $c_{p,l}$ are the density and specific heat capacity of the droplets assumed to consist of water. $\nu_{air}$ is the kinematic viscosity of air. $k_g$ is the thermal conductivity of gas which relates to the thermal diffusivity of gas $D_g$ by $k_g=D_g \rho_g c_{p,g}$. $\rho_g$ and $c_{p,g}$ are the density and specific heat capacity of gas, Note that $\rho_g c_{p,g}=(\rho_a c_{p,a} + \rho_v c_{p,v})$, with $\rho_a$ and $\rho_v$ being the densities of air and vapour and $c_{p,a}$ and $c_{p,v}$ being the specific heat capacities of air and vapour. $c_{drop}$ and $c_{fluid}$ denote the vapour mass fractions of the droplet and the fluid at the location of the droplet. $\beta$ is the density ratio term defined as $\beta=3\rho_g/(\rho_g+2\rho_l)-1$.

\Equrefs{eqn:DropletTemp} and \equrefs{eqn:DropletDiam} are closed using the Ranz-Marshall correlations \cite{ranz1952Part2}, which are reasonable assumptions for moving point droplets in our study since the droplet Reynolds number are at most O($100$), in accordance with the applicable limits \cite{ellendt2018model}. The correlations give us the estimation of $h_m$ and $\Sher_{drop}$ for a single spherical droplet.
\begin{equation}
    \Sher_{drop} = 2 + 0.6 \Rey_{drop}^{1/2}(\nu_{air}/D_{vap})^{1/3}, \label{eqn:ShRanzMarshall}
\end{equation}
\begin{equation}
    h_m r/(D_gc_{p,g}\rho_g) = 2 + 0.6 \Rey_{drop}^{1/2}(\nu_{air}/D_{g})^{1/3}, \label{eqn:NuRanzMarshall}
\end{equation}
where we have droplet Reynolds number
\begin{equation}
    \Rey_{drop} = \frac{|u_{i,g,n}-u_{i,n}| (2r)}{\nu_{air}}.
\end{equation}
Note that, by definition, $\Sher_{drop}$ and $\Rey_{drop}$ are also functions of $r$. The realistic parameters needed are listed in \tabrefs{tab:realpara}.

The relative humidity at the location of the droplets is defined as $P_{vap,drop}/P_{sat} = c/c_{sat,vap}$, where the vapour mass fraction $c$ is solved from \equrefs{eqn:GasMass}. Therefore, to calculate the local relative humidity, the saturated vapour mass fraction $c_{sat,vap}$ is first determined by the ideal gas law:
\begin{equation}
   c_{sat,vap}=\frac{P_{sat}}{\rho_g R\theta_g},
\end{equation}
where $P_{sat}$ and $R$ are the saturated vapour pressure, and specific gas constant of water vapour. The saturated pressure can be obtained through the Antoine's relation as
\begin{equation}
    P_{sat}\left(\theta_g\right)=10^5\textrm{exp}\left(11.6834-\frac{3816.44}{226.87+\theta_g-273.15}\right).
\end{equation}

\begin{table}[ht]
\begin{center}
\begin{tabular}{|c|c|c|}
\hline
Cough properties & Symbols & Values\\
\hline
Temperature of vapour puff & $\theta_{cough}$ & $\SI{34}{\degreeCelsius}$ \\
Mean velocity of cough & $U_{cough}$ & $\SI[per-mode=repeated-symbol]{11.2}{\meter \per \second}$ \\
Diameter of mouth & $D_{mouth}$ &  $\SI{2.3}{\centi \meter}$ \\
\hline
Properties for gas phase & Symbols &Values \\
\hline
Ambient temperature & $ \theta_{ambient}$ & $\SI{20}{\degreeCelsius}$ \\
Diffusivity of water vapour in the gas phase & $D_{vap}$ & $2.5\times 10^{-5} \si[per-mode=repeated-symbol]{\square \meter \per \second}$ \\
Thermal diffusivity of gas & $D_g$ &  $2.0 \times 10^{-5} \si[per-mode=repeated-symbol]{\square \meter \per \second}$ for air at $\SI{25}{\degreeCelsius}$ \\
Air kinematic viscosity & $\nu_{air}$ & $1.562 \times 10^{-5} \si[per-mode=repeated-symbol]{\square \meter \per \second}$ \\
Density of the gas & $\rho_{g}$ & $\SI[per-mode=repeated-symbol]{1.204}{\kilogram \per \cubic \meter}$ \\
Thermal expansion coefficient & $ \beta_\theta$ & $3.5\times 10^{-3} \si[per-mode=repeated-symbol]{\per \kelvin}$ \\
Expansion coefficient for vapour mass fraction& $ \beta_c$ & $0.5 \si[per-mode=repeated-symbol]{\cubic \meter \per \kilogram} \times 1.204 kg/m^3 \approx 0.6 $ \\
Specific heat capacity of gas & $c_{p,g}$ & $\SI[per-mode=repeated-symbol]{1}{\kilo \joule \per \kilogram \per \kelvin}$ \\
Specific gas constant of water vapour & $R$ & $461.5  \si[per-mode=repeated-symbol]{\joule \per \kilogram \per \kelvin}$\\ 
\hline
Properties for droplets & Symbols & Values\\
\hline
Droplet density & $\rho_{l}$ & $\SI[per-mode=repeated-symbol]{993}{\kilogram \per \cubic \meter}$ \\
Particle density parameter & $\beta$ & $ {-0.9982}$\\
Specific heat capacity of droplet & $c_{p,l}$ & $\SI[per-mode=repeated-symbol]{4.186}{\kilo \joule \per \kilogram \per \kelvin}$ \\
Latent heat of water & $L$ & $\SI[per-mode=repeated-symbol]{2256}{\kilo \joule \per \kilogram}$\\
\hline
\end{tabular}
\caption{Definition of realistic parameters employed in the numerical simulations of this study. Droplets are assumed to contain pure water.}
\label{tab:realpara}
\end{center}
\end{table}

\section*{Coupling methodology}
To interpolate gas quantities to the droplet locations, we employ the tri-cubic Hermitian interpolation scheme, which are sufficiently accurate for turbulent flows and comparable to B-spline interpolations \cite{van2013optimal,ostilla2015multiple}. The backwards forcing of the droplets onto the gas phase uses the tri-linear projection onto the eight nearest nodes to the droplets location \cite{mazzitelli2003relevance}. 

\section*{Numerical setup}
To numerically solve the equations, we used our finite difference solver AFiD \cite{vanderpoel2015} with high-performance Message Passing Interface (MPI) and point-particle model. The size of the computational domain in dimensional form is $\SI{0.37}{\meter}$ (spanwise length) $\times \SI{0.37}{\meter}$ (height) $\times \SI{1.10}{\meter}$ (streamwise length) and is tested to be large enough to capture the cough vapour and spreading droplets. The grid points chosen are $512\times 512\times 1536$ to ensure that enough resolution has been employed. The mouth is modelled as a circular inlet centred at mid-height of the domain.

\section*{Initial droplet distribution}
The topic of distribution of the initial droplet sizes  is in itself a subject of considerable importance and active debate \cite{duguid1946,xie2009exhaled,johnson2011,somsen2020}. Here, we seeded the respiratory event with droplets with initial diameters ranging between \SI{10}{\micro \meter} and roughly \SI{1000}{\micro \meter}, based on an experimental measurement \cite{xie2009exhaled}. Estimates of the droplet volume fractions give O($10^{-5}$) and so droplets are assumed not to collide or coalesce. These droplets also do not set the relative humidity of the puff, since the relative humidity of the puff at the inflow is assumed to be 100\%. Note that even smaller droplets could be added, but they would further extend the required CPU time and are not necessary to convey the main message of this work.

\section*{Cough properties}
The cough temporal profile we apply is the average of the spirometer measurements shown in figure 2 in the manuscript. We did not impose any noise on the inflow velocity, and the transition to turbulence was not forced, but occurred naturally in the simulation. The inlet velocity, temperature and vapour mass fraction spatial profiles are prescribed as a Gaussian profile centred at the mouth location. Note that we set saturated vapour mass fraction at the inlet, which is calculated based on temperature of the human respiratory tract \cite{morawska2009size}. For the initial droplet size, we employ a similar distribution as \cite{xie2009exhaled} with $2500$ droplets. The droplets are randomly positioned at the inlet and evenly injected in time with the velocity matching the local inlet velocity.

\section*{Initial and boundary conditions}

The flow field is initialised as quiescent with uniform atmospheric pressure, temperature (20$^{\circ}$C) and a relative humidity of 50\%. Neumann condition is applied to the boundary of the simulation domain, except for the mouth opening where the injection velocity is given. When a droplet reaches the boundary of the domain, it is flagged as `dead' and then removed from the simulation. The simulation of a droplet also terminates when its diameter falls below 4.6$\upmu$m, as simulating smaller droplets require significantly more CPU time without altering the main results.

\bibliographystyle{elsarticle-num}
\bibliography{literature}